\documentstyle[seceq,mbf,wrapft,epsf]{ptptex}

\def\lsim{\mathop{\rlap{\raisebox{0.6ex}{$<$}}\raisebox{-0.6ex}{$\sim$}}}
\def\gsim{\mathop{\rlap{\raisebox{0.6ex}{$>$}}\raisebox{-0.6ex}{$\sim$}}}

\makeatletter
\def\maketitle{
\vspace*{-20mm}
\leftline{\epsfbox{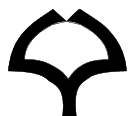}}
\vspace{-10.0mm} 
{\baselineskip-4pt
\font\yitp=cmmib10 scaled\magstep2
\font\elevenmib=cmmib10 scaled\magstep1  \skewchar\elevenmib='177
\leftline{\baselineskip20pt
\hspace{10mm} 
\vbox to0pt
   { {\yitp\hbox{Osaka \hspace{1.5mm} University} }
     {\large\sl\hbox{{Theoretical Astrophysics}} }\vss}}
\rightline{\large\baselineskip20pt\rm\vbox to20pt{
\baselineskip14pt
\hbox{OU-TAP-192}
\vspace{1mm}
\hbox{\today}\vss}}%
}
\par
\vspace*{5mm}
 \begingroup
 \if@twocolumn
 \twocolumn[\@maketitle]
 \else \relax
 \global\@topnum\z@ \@maketitle \fi\thispagestyle{empty}\@thanks
 \endgroup
 \if 0\@prtstyle \relax 
 \fi
 \let\maketitle\relax
 \let\@maketitle\relax
 \gdef\@thanks{}\gdef\@author{}\gdef\@title{}\let\thanks\relax}
\def\@maketitle{
\null \vskip -13.45pt
\def\hang{\par\hangindent=0pt\hangafter=1\noindent}
\@ifundefined{@addenda}{%
\@ifundefined{@supple}{
\centerline{\footnotesize%
\@ifundefined{@pubinfo}{~~}{%
\if 2\@prtstyle Prog.~Theor.~Phys. \@pubinfo, Letters
\else Progress of Theoretical Physics, \@pubinfo
\fi \gdef\@pubinfo{}}}
\par \vskip 12.3pt}%
{\vskip 18pt}
\begin{center}
{\large\bf \hang \@title}
\@ifundefined{@subtitle}{}
{\par\vskip 5pt {\large\it\hang \rule[.5ex]{1.7em}{\hruleheight}\ %
\@subtitle\ \rule[.5ex]{1.7em}{\hruleheight}} \gdef\@subtitle{}}
\par\vskip 12.3pt
{\hang \@ifundefined{@author}{}{\@author}} \par\vskip 7.47pt
\@ifundefined{@inst}{}%
{\instsize\it \hang \@inst \gdef\@inst{}\par \vskip 11pt}
\end{center}
\@ifundefined{@recdate}{}%
{\hang{\footnotesize \begin{center}(Received \@recdate{})\end{center}}
\gdef\@recdate{}\par \vskip 5.75pt}
\@ifundefined{@abst}{}%
{\begingroup
\begin{center}
\parbox[t]{\@abstwidth}{\setlength{\parindent}{1.4em}%
\abstsize \@abst}
\gdef\@abst{} \par
\end{center}
\endgroup
\vskip20pt}}%
{\centerline{\Large\bf \@addenda}
\par \vskip 20pt
\begin{center}
{\bf \hang \@title}
\@ifundefined{@subtitle}{}
{\par \vskip 2.5pt {\it\hang \rule[.5ex]{1.7em}{\hruleheight}\ %
\@subtitle\ \rule[.5ex]{1.7em}{\hruleheight}} \gdef\@subtitle{}}
\par\vskip 10pt
{\hang \@author} \par\vskip 3pt
\@ifundefined{@publishedin}%
{\immediate\write20{>Warning: <publishedin> is NOT defined!!}
Write ``Prog.~Theor.~Phys.~Vol.~(Year), Page'' in 
{\tt $\backslash$publishedin}!!}
{{\hang \@publishedin}\gdef\@publishedin{}}
\par\vskip 15pt
\@ifundefined{@recdate}{}
{{\hang (Received \@recdate{})}
\gdef\@recdate{}\par \vskip 5pt}
\end{center}}
}
\def\invitedpaper{\gdef\@invited{INVITED PAPERS}}
\def\ps@headings{
\let\@mkboth\markboth
\if 0\@prtstyle
\def\@oddhead{}
\def\@oddfoot{\hbox to \textwidth{\hfil\rm\thepage\hfil%
\makebox[0pt][r]{\@typeset}}}
\else
\def\@oddfoot{\hfil\@typeset}
\@ifundefined{@supple}{%
\def\@oddhead{\@ifundefined{@invited}{}{%
\fbox{\rule[-.7pt]{0pt}{.84em}\hspace{.5pt}\@invited}}
\hfil\rm\thepage}
}{
\def\@oddhead{\hbox to \textwidth{\hfil{\footnotesize%
\@supple\@ifundefined{@pubinfo}{}{\@pubinfo}}%
\hfil\makebox[0pt][r]{\rm\thepage}}}
\def\@evenhead{\hbox to \textwidth{\makebox[0pt][l]{\rm\thepage}%
\hfil{\footnotesize\@supple\@ifundefined{@pubinfo}{}{\@pubinfo}}%
\hfil}}}
\fi}
\makeatother

\title{
Braneworld inflation driven by dynamics of a bulk scalar field
}

\author{ 
Yoshiaki
{\sc Himemoto}\footnote{E-mail:himemoto@vega.ess.sci.osaka-u.ac.jp}
and Misao
{\sc Sasaki}\footnote{E-mail: misao@vega.ess.sci.osaka-u.ac.jp}
}

\inst{
 Department of Earth and Space Science, Graduate
School
of Science,\\
Osaka University, Toyonaka 560-0043, Japan}

\abst{We review a viable alternative scenario of the inflationary
universe in the context of the Randall-Sundrum (RS) braneworld.
In this scenario, the dynamics of a 5-dimensional scalar
 field, which we call a bulk scalar field, plays the central role.
Focusing on the second (single-brane) RS model, we discuss
braneworld inflation driven by a bulk scalar field without introducing
 an inflaton on the brane.
As a toy model, for the bulk scalar field, we
 consider a minimally coupled massive scalar field in the 5-dimensional
 spacetime, and look for a perturbative solution of the field equation
 in the anti-de Sitter background with an inflating brane. For a suitable
range of the model parameters, we
 find a solution that realizes slow-roll inflation on the brane. When
 the Hubble parameter on the brane and the mass of a bulk scalar field
 are much smaller than a typical 5-dimensional mass scale, it is found
 that this proposed inflation scenario reproduces the standard inflation
 scenario in the 4-dimensional theory.}

\begin{document}

\maketitle

\section{Introduction}
It is very likely that our four-dimensional universe
is a subspace of a higher-dimensional spacetime.
In fact, string theory, which is a candidate for the unified theory,
is a higher-dimensional theory.
Therefore it is of great interest to develop a higher dimensional
cosmological scenario which is consistent with existing observational
data and which predicts new phenomena that can be experimentally
or observationally tested. In particular,
the braneworld scenario has attracted much attention
\cite{braneworld}
and a model proposed by Randall and
Sundrum (RS2)\cite{Randall:1999vf} ignited active research
of braneworld cosmology.\cite{cosmology}

As an alternative to the standard 4-dimensional theory of inflation,
we recently proposed a brane cosmological model
in which slow-roll inflation is driven not by an inflaton on the
brane but solely by dynamics of a scalar field living
in the 5-dimensional bulk.\cite{Himemoto:2001nd}
The scalar field we introduced in Ref.~\citen{Himemoto:2001nd}
 should probably be a dilaton-like gravitational field from
the unified theoretical point of view.
In fact, it is natural that the 5-dimensional theory is itself an
effective theory which originates from a yet higher-dimensional
theory, and the 5-dimensional effective action includes
some scalar fields of gravitational origin.
Provided there is a bulk scalar field with a suitable potential,
it was shown in Ref.~\citen{Himemoto:2001nd}
that there exists a field configuration in the bulk
that indeed realizes inflation on the brane.
In particular,
we found that the standard inflationary cosmology is
reproduced, when $|m^{2}|\ell^{2}\ll 1$
and $H^{2}\ell^{2}\ll 1$ are satisfied,
where $\ell$ is the curvature radius of AdS$_5$,
$m$ is the mass of the bulk scalar
field, and $H$ is the Hubble parameter on the brane.
\cite{Kobayashi:2001yh,Sago:2002gi,Yokoyama:2001nw}

This paper is organized as follows.
In \S2, we review the basic picture of the inflation scenario driven by
a bulk scalar field. In \S3,
we investigate the behavior of the bulk scalar field
on the inflating braneworld
and discuss the effective description of
the dynamics from the 4-dimensional point of view.
Section 4 is devoted to conclusion.

\section{Braneworld inflation without inflaton on the brane}

Based on the braneworld scenario of the RS2 type,
we consider a 5-dimensional bulk
with a single positive tension brane which is located
at the fixed point of the $Z_2$ symmetry.
We assume that the 5-dimensional gravitational equations are
\begin{equation}
  R_{ab}-{1\over2}g_{ab}R+\Lambda_5g_{ab}
=\kappa_5^2\left(T_{ab}+S_{a b} \delta(r-r_0)\right)\,,
\label{bulkgreq}
\end{equation}
where $r$ is the coordinate normal to the brane and
the brane is assumed to be located at $r=r_0$.
As for $S_{ab}$ and $T_{ab}$,
we neglect the contribution to $S_{ab}$ from
matter fields confined on the brane
and consider a minimally coupled bulk scalar field
with the potential $V(\phi)$.
Thus we have
\begin{eqnarray}
S_{a b} = -\sigma q_{a b}\,, \quad {\mathrm{and}} \quad
T_{a b} = \phi_{,a} \phi_{,b}- g_{a b}\left( {1\over2}
g^{c d}\phi_{,c} \phi_{,d}+ V(\phi)\right),
\label{5emt}
\end{eqnarray}
where $\sigma$ is the tension of the brane and $q_{ab}$ is the induced
metric on the brane.
In order to recover the Randall-Sundrum flat braneworld
when $T_{ab}$ vanishes,
we choose $\Lambda_5$ as
$\Lambda_5=-\kappa_5^4\sigma^2 / 6$.
Then, the effective 4-dimensional Einstein equations on the brane
become
\cite{Himemoto:2001nd,Shiromizu:2000wj}
\begin{equation}
G_{\mu \nu}=
\kappa_{4}^{2}T_{\mu \nu}^{(s)}-E_{\mu \nu},
\label{Einstein}
\end{equation}
where
\begin{eqnarray}
\kappa_{4}^{2}&=&{\kappa_{5}^{4}\sigma \over 6}
={\kappa_5^2\over\ell_0}, \\
T_{\mu \nu}^{(s)} & = &{\ell_0\over6}
\left(4 \phi_{,\mu} \phi_{,\nu}+\left({3\over2}(\phi_{,r})^{2}
-{5\over2}q^{\alpha \beta}\phi_{,\alpha} \phi_{,\beta}
-3V(\phi)\right)
q_{\mu \nu}\right),\\
E_{\mu \nu}&=& \,{}^{(5)}C_{r b r d}
\,q_{\mu}^{b}\,q_{\nu}^{d}\,.
\end{eqnarray}
Here $\ell_0=6/(\kappa_5^2\sigma)$ is the AdS$_5$
curvature radius and ${}^{(5)}C_{r b r d}$
is the $5$-dimensional Weyl tensor with its two indices
projected in the $r$-direction.

Focusing on the zeroth order description of the cosmological model,
we consider the case in which the metric induced on the
brane is isotropic and homogeneous.
Because of the assumed $Z_2$ symmetry, the boundary condition for
the bulk scalar field at the position
of the brane is given by \footnote{For simplicity, we do not consider a
possible coupling of $\phi$ to the metric on the brane, though an
extension to such a case may be worth investigating in the future.}
\begin{equation}
\partial_{r}\phi|_{r=r_0}=0.
\label{bc}
\end{equation}
Then, the 4-dimensional effective Friedmann equation is given by
\begin{equation}
3\left[\left({\dot a\over a}\right)^{2}+{K\over a^2}\right]
\equiv 3H^{2}
=\kappa_4^{2} \rho_{\mathrm{eff}}\,,
\label{friedmann}
\end{equation}
with
\begin{eqnarray}
\rho_{\mathrm{eff}}={\ell_0\over2}
\left({\dot{\phi}^{2}\over2}+V(\phi)\right)
-{\ell_0\over\kappa_5^2}E_{tt}\,,
\label{rhoeff}
\end{eqnarray}
where $a(t)$ is the cosmological scale factor of the brane
and $K=\pm1$, $0$.
The equations for $\phi$ and $E_{tt}$,
are basically 5-dimensional.
However, in the present spatially homogeneous case,
the Bianchi identities supply the evolution equation of
$E_{tt}$ on the brane as\cite{Himemoto:2001nd}
\begin{equation}
E_{tt}={\kappa_5^{2}\over{2a^{4}}}\int^t a^{4}\dot{\phi}
(\partial_{r}^{2}\phi + {\dot a\over a}\dot{\phi})\,dt.
\label{weyl}
\end{equation}
We then see that $E_{tt}$ can be neglected if both $\dot{\phi}$ and
$\partial_{r}^{2}\phi$ are sufficiently small on the brane.
Thus a sufficient condition for inflation to occur on the brane is that
$\phi$ is a slowly varying function with respect to both $t$ and $r$ in
the vicinity of the brane.

\subsection{The field equation in the de Sitter brane background}
Our purpose is to find a solution of the field equations that has
nontrivial dynamics in the bulk and gives rise to inflation on the
brane. 	For this purpose,
we investigate the general behavior of the solution of 5-dimensional scalar
field equation in AdS$_5$ background with an inflating brane.

As a toy model, we assume the potential of the form,
\begin{equation}
V(\phi)=V_0+{1\over2}m^{2}\phi^{2},
\label{potential}
\end{equation}
and consider the region $|m^2|\phi^2/2\ll V_0$.
Here we do not specify the signature of $m^{2}$, though
$m^2$ should be negative for inflation to end if the model
should describe the braneworld inflation self-consistently.
Then, the effective 5-dimensional cosmological constant
becomes
\begin{equation}
\Lambda_{5,\rm{eff}} =
\Lambda_{5} + \kappa_5^2V_0\,.
\label{eq:Lambda5eff}
\end{equation}
Here we assume that $|\Lambda_5|>\kappa_5^2V_0$ so that the background
spacetime is still effectively AdS$_5$ with the effective curvature
radius
 $\ell = \left|{6}/{\Lambda_{5,{\rm eff}}} \right|^{1/2}$.
Note that $\ell>\ell_0$ where $\ell_0=|6/\Lambda_5|^{1/2}$.
Then the bulk metric may be written as \cite{Garriga:2000bq}
\begin{eqnarray}
ds^2&=&dr^2+(H\ell)^2\sinh^2(r/\ell)\,
[-dt^2+H^{-2}e^{2Ht}d \mbox{\boldmath $x$}^2]\,.
\label{eq:bulkmetric}
\end{eqnarray}
Here we have adopted the spatially flat slicing ($K=0$)
of de Sitter space for simplicity.
Note that the Friedmann equation~(\ref{friedmann}) at the lowest
order determines the Hubble parameter $H$ as
$H^2={\kappa_5^2V_0}/{6}$.

At the first order in the amplitude of $\phi$, the background
metric is unaffected. Hence we look for a perturbative
solution for $\phi$ on this effective AdS$_5$ background.
Then the field equation in the bulk becomes
\begin{equation}
(-\Box_{5}+m^{2})\phi=
  \left[-\hat L_r +{1
    \over (H\ell)^2 \sinh^2(r/\ell)}\left(
  \hat L_t-H^2 e^{-2Ht}\partial_{\mbox{\boldmath $x$}}^2 \right)\right]
  \phi=0\,,
\label{fieldeq}
\end{equation}
where
\begin{equation}
\hat L_t={\partial^2\over \partial t^2}+3H{\partial\over\partial t}
\quad {\mathrm{and}} \quad
\hat L_r={1\over \sinh^4(r/\ell)}{\partial\over\partial r}
\sinh^4(r/\ell){\partial\over\partial r} -m^2,
\label{spaceeq}
\end{equation}
with the boundary condition (\ref{bc}).
The equation of motion for the $\mbox{\boldmath $x$}-$independent
bulk scalar field can be separated by setting $\phi=u(r)\psi(t)$
as
\begin{equation}
\left[\hat L_r+{\lambda^{2}\over{\ell^{2}}
\sinh^{2}(r/\ell)}\right] u(r)=0\,,
\label{doukeieq}
\end{equation}
\begin{equation}
\left[\hat L_t+\lambda^{2}H^{2}\right]
\psi(t)=0\,.
\label{timeeq}
\end{equation}
Eq.~(\ref{doukeieq}) determines the spectrum of $\lambda^{2}$,
while Eq.~(\ref{timeeq}) is the equation of motion
for a 4-dimensional scalar field with effective mass-squared
$\lambda^{2}H^{2}$ in a de Sitter space of radius $H^{-1}$.

To see the structure of the mass spectrum,
it is convenient to rewrite Eq.~(\ref{doukeieq}) in the
standard Schr\"odinger
form. To do so, we introduce the conformal radial coordinate $y$
through $dr/R(r)=dy$,
where $R(r)=\ell\sinh(r/\ell)$.
Then the metric (\ref{eq:bulkmetric}) is expressed as
\begin{equation}
ds^{2}=R^{2}
\left(dy^{2}-H^{2}dt^{2}+e^{2Ht}\,d\mbox{\boldmath $x$}^{2}\right),\quad
 (-\infty<y<+\infty)
\end{equation}
where
$R(y)=\ell\sinh^{-1}(|y|+y_0)$
and $y_0$ is defined by
$\sinh(y_0)=\sinh^{-1}(r_0/\ell)=H\ell$.
Then putting $u=R^{-3/2} f(y)$,
Eq.~(\ref{doukeieq}) becomes
\begin{equation}
-f''+\tilde{V}f=\lambda^{2}f\,,
\label{syure}
\end{equation}
where the prime denotes the $y$-derivative and
$\tilde{V}$ takes the form of a ``volcano potential'',
\begin{eqnarray}
\tilde{V}
= {9\over4}+{{15+4m^{2}\ell^{2}}\over 4 \sinh^{2}(|y|+y_0)}
-3\coth(|y|+y_0) \delta(y).
\end{eqnarray}
It is clear that the volcano potential $\tilde{V}$ approaches
the constant 9/4 at $y \rightarrow \pm \infty$.
We find that Eq.~(\ref{syure}) has a normalizable
bound-state solution in the region
\begin{equation}
\lambda^{2}<9/4 \,,
\end{equation}
for $m^{2}$ smaller than a critical value
($\sim 9H^2/2$) and the continuous spectrum starts at
\begin{equation}
\lambda = 3/2.
\end{equation}
Here we note that this bound-state solution
corresponds to the zero-mode solution $u=$ const in the case $m^{2}=0$.
We also note that the continuous spectrum is
independent of the 5-dimensional mass of the field.
These continuous modes are called the Kaluza-Klein
modes and the existence of them is the main signature of the braneworld.

Denoting the bound state mode simply by $\psi(t)u(r)$
and a Kaluza-Klein mode by $\psi_p(t)u_p(r)$
where $p^2=\lambda^2-9/4$,
the general solution for $\phi$ is given by
\begin{eqnarray}
\phi=\psi(t)u(r)
+\int_{-\infty}^\infty dp\, \psi_p(t)u_p(r)\,.
\label{general}
\end{eqnarray}
It should be noted that the mode functions $u(r)$ and $u_p(r)$,
which satisfy the boundary condition (\ref{bc}), are
square-integrable in the usual sense but are singular
in the limit $r\to0$ except for $u(r)$ in the case of
negative $m^2$.

\section{Bulk scalar field mimicking 4d inflaton dynamics}

In Ref.~\citen{Himemoto:2001nd}, we assumed $m^2<0$ and
that the scalar field is described by the zero-mode solution
$\phi=\psi(t)u(r)$.
Then it was shown that the dynamics of this system
may be well described by the effective 4-dimensional scalar field
and the standard slow-roll inflation is realized
by the bulk scalar field
when $|m^{2}|\ell^{2}\ll 1$ and $H^{2}\ell^{2} \ll 1$.

However, as given by Eq.~(\ref{general}), the general solution
will contain the contribution from the Kaluza-Klein modes.
Furthermore, if the backreaction of the scalar field to
the geometry is taken into account, which appears at
the second order in $\phi$, the inclusion of the Kaluza-Klein
modes is indespensable to make the geometry at $r=0$ (for
fixed $t$) regular even in the case $m^2<0$.\footnote{Although
$u(r)$ vanishes at $r=0$ for $m^2<0$, its derivative diverges
for $|m^2|\lsim H^2$.}
Therefore, for the scenario to be viable, it is necessary to
show that the inclusion of the Kaluza-Klein modes does not
affect the dynamics of the brane too much.
In this section, we analyze the general behavior of
the scalar field for a much more general class of initial conditions
for which there is no need to assume
the separable form nor the case $m^2<0$.
Then we show how the description in terms of the effective
4-dimensional (zero mode) field is recovered on the brane.

\subsection{Dynamics induced on the inflating brane}

We consider arbitrary, regular initial data for this
scalar field and investigate the generic behavior of the scalar
field at sufficiently late times by
analyzing the properties of the retarded Green function.

We start with the construction of the Green function.
The retarded Green function satisfies
\begin{equation}
(-\Box_{5}+m^{2})G(x,x')={\delta^{5}(x-x')\over \sqrt{-g}}\,,
\label{green}
\end{equation}
with the causal condition that $G(x,x')=0$ for $x'$ not in the
causal past of $x$.
For given initial data on the hypersurface $t=t_i$,
the time evolution of a scalar field is given by
\begin{eqnarray}
 \phi(x)
= \int_{t'=t_i}
           [(N^{a}\partial'_{a} G(x,x'))\phi(x')
-G(x,x') N^{a}\partial'_{a} \phi(x')]
            \sqrt{\gamma(x')}\  d^{4}x',
\end{eqnarray}
where $N^{a}$ is the time-like unit vector normal
to the initial hypersurface,
and $\gamma$ is the determinant of the metric induced on
this initial hypersurface.
Since we are interested in the spatially homogeneous brane,
we focus on the $\mbox{\boldmath $x$}$-independent scalar field
configurations.
Namely, we consider the spatially averaged Green function
defined by
\begin{equation}
{\cal G}(t,r;t',r'):=\int d{\mbox{\boldmath $x$}}'\, G(x,x').
\end{equation}
Since $G(x,x')$ depends on the spatial
coordinates ${\mbox{\boldmath $x$}}$ and ${\mbox{\boldmath $x$}}'$ though
the form $|{\mbox{\boldmath $x$}}-{\mbox{\boldmath $x$}}'|$,
the ${\mbox{\boldmath $x$}}$-dependence also
disappears after taking the average over ${\mbox{\boldmath $x$}}'$.
The equation for ${\cal G}(t,r;t',r')$ follows
from Eq.~(\ref{green}) as
\begin{equation}
  \left[{\hat L_t\over (H\ell)^2\sinh^2(r/\ell)}-\hat L_r
   \right]{\cal G}(t,r;t',r')={\delta(t-t')\delta(r-r')
    \over H\ell^4\sinh^4(r/\ell)e^{3Ht}}\,,
\label{green2}
\end{equation}
where $\hat L_t$ and $\hat L_r$ are the operators defined in
Eq.~(\ref{spaceeq}).

We briefly explain how to construct the Green function
${\cal G}(t,r;t',r')$(see Ref.~\citen{Himemoto:2001hu} for details).
We begin by considering a set of eigenfunctions $\psi_p(t)$ of
the operator $\hat L_t$.
The equation is given by Eq.~(\ref{timeeq}),
with the change of labeling from $\lambda$ to $p$
($\lambda^2=p^2+9/4$).
We find
\begin{eqnarray}
\psi_p(t)  = (2\pi)^{-1/2} e^{(-ip-3/2)Ht}\,.
\end{eqnarray}
Recall that $(p^2+9/4)H^2$ is the four-dimensional effective
mass squared for each mode.
The functions $\psi_p(t)$ satisfy the orthonormality
and the completeness conditions for real $p$.

Next we consider the eigenfunctions $u_p(r)$ for $\hat L_r$.
The equation for $u_p(r)$ is given by Eq.~(\ref{doukeieq}).
We denote an eigenfunction which is regular
on the upper half complex $p$ plane
by $u_p^{\mathrm{(out)}}(r)$.
It is given by
\begin{eqnarray}
 u_p^{\mathrm{(out)}}(r) & = &
  {\Gamma(1-ip)\over2^{ip}}{P_{\nu-1/2}^{ip}(\cosh (r/\ell))
  \over{\sinh^{3/2}(r/\ell)}}
    \displaystyle\mathop{\sim}_{r\to 0}  (r/\ell)^{-ip -3/2},
\end{eqnarray}
where $P_{\nu-1/2}^{ip}(z)$ is the associated Legendre function of the
first kind and
$\nu=\sqrt{m^2\ell^2+4}$.
The reason for assigning the superscript `(out)' to this eigenfunction
is that $\psi(t) u_p^{\mathrm{(out)}}(r)\propto e^{-ip(Ht+\ln r)}$
describes a wave propagating out to the Cauchy horizon
given by $r\to0$ with $Ht+\ln r=\mbox{const}$.
On the other hand,
we denote the eigenfunction which satisfies the Neumann boundary condition
at the position of the brane (\ref{bc}) by $u_p^{(Z_2)}(r)$.
Here the superscript `$(Z_2)$' is assigned because of its $Z_2$-symmetric
property.
We can describe $u_p^{(Z_2)}(r)$ by a linear combination
of two independent solutions as
$ u_p^{(Z_2)}(r)=u_p^{\mathrm{(out)}}(r)-\gamma_p
u_{-p}^{\mathrm{(out)}}(r)$ and
where $\gamma_{p}$ is determined by the Neumann boundary condition at the brane.

With these eigenfunctions, we can
express the Green function as
\begin{equation}
 {\cal G}(t,r;t',r')
   =\int_{\cal C}dp \,{\cal G}_p(r,r')\psi_p(t)\psi_{-p}(t'),
\label{calG2}
\end{equation}
where ${\cal C}$ is a path extending from $p=-\infty$ to
$p=\infty$ on the complex $p$-plane with the property that
the integrand contains no pole nor branch cut above the path
(see Fig.~1),
and ${\cal G}_p$ is constructed from $u_p^{\mathrm{(out)}}$
and $u_p^{(Z_2)}(r)$ as
\begin{equation}
  {\cal G}_p(r,r')={1\over W_p}
    \left(u_p^{\mathrm{(out)}}(r) u_p^{(Z_2)}(r') \theta(r'-r)
          +u_p^{\mathrm{(out)}}(r') u_p^{(Z_2)}(r) \theta(r-r')\right),
\label{gp}
\end{equation}
with $W_p$ being the Wronskian given by
\begin{eqnarray}
 W_p & =& \ell^4 \sinh^4(r/\ell)
       \left[ (\partial_r u_p^{\mathrm{(out)}}(r))  u_p^{(Z_2)}(r)
              -u_p^{\mathrm{(out)}}(r) (\partial_r u_p^{(Z_2)}(r))\right]\cr
     & = & 2ip \ell^3 \gamma_p.
\label{wronskian}
\end{eqnarray}
\begin{center}
\leavevmode
\epsfysize=8cm\epsfbox{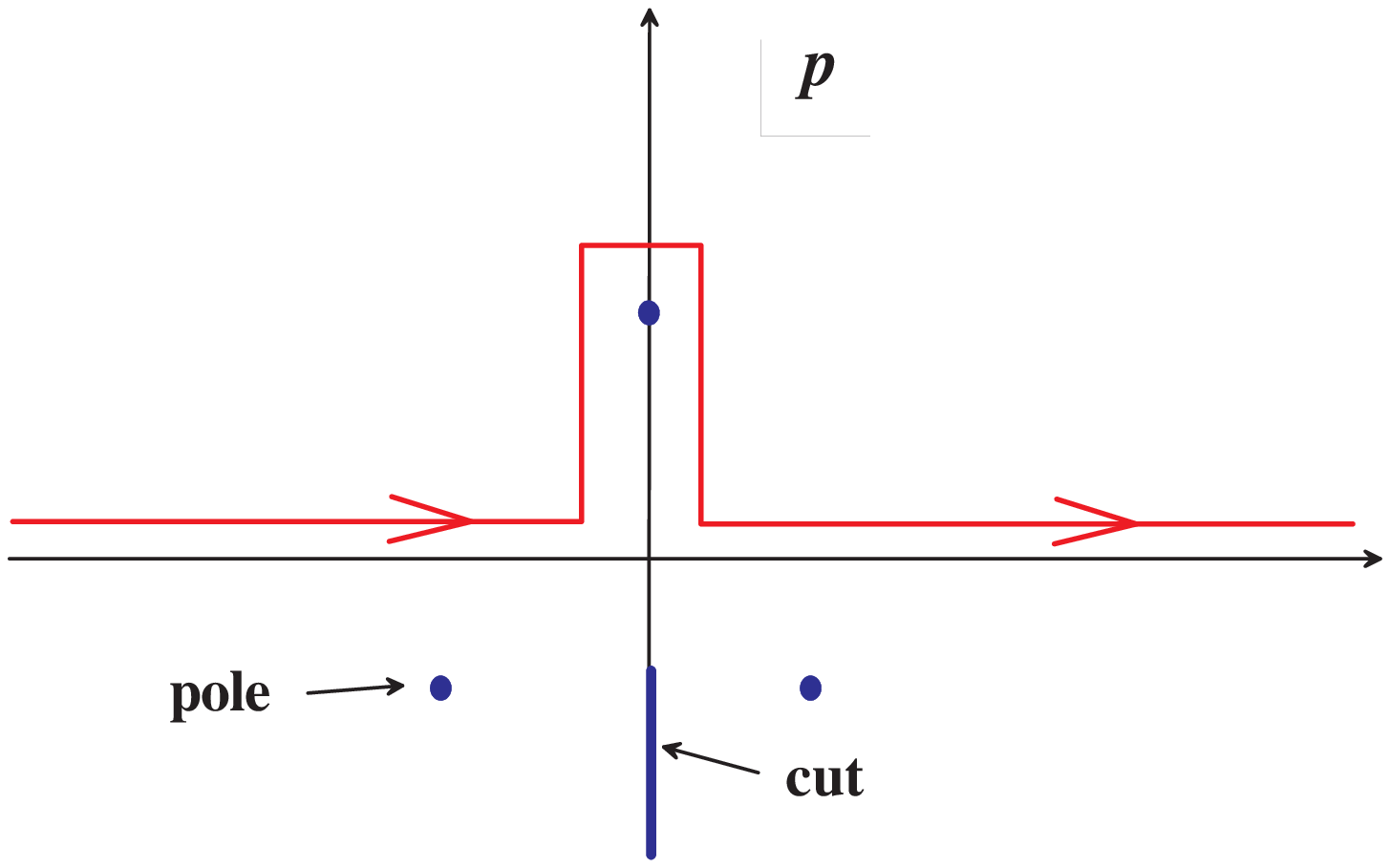}
\\
Fig. 1. Contour of integration for the retarded Green function.
\end{center}
\vspace{5mm}

With the above choice of the complex path ${\cal C}$,
the retarded boundary condition is satisfied, and the Green function
vanishes for spatially separated $(t,r)$ and $(t',r')$.
In particular, it vanishes in the limit $r\to 0$ for fixed values
of $t$ and $t'$. This guarantees the regularity at $r=0$ of
the scalar field for arbitrary, regular initial data.
As a result, the late-time behavior is understood by
investigating the structure of
singularities such as poles and branch cuts
in the Green function ${\cal G}_p$.
The singularity on the complex $p$ plane with
the largest imaginary part
dominates the late-time behavior.

For $H^{2}\ell^{2}\ll 1$ and $m^{2}\ell^{2}\ll 1$,
the equation that determines the location of a pole
with the smallest imaginary part is given
by\cite{Himemoto:2001hu}
\begin{equation}
{m^{2}\ell^{2}\over 2}
-\left(p^{2}+{9\over4}\right)H^{2}\ell^{2} \approx 0.
\label{leading}
\end{equation}
This equation is always a good approximation irrespective of the value of
$m^{2}/H^{2}$. We denote the two solutions by
\begin{equation}
 p_{\pm} =\pm i\sqrt{{9\over 4}-{m_{\mathrm{eff}}^2\over H^2}}
+O\left((H\ell)^2,(m\ell)^2\right),
\label{ppm}
\end{equation}
where
\begin{equation}
m_{\mathrm{eff}}^{2}
={m^{2}\over 2}.
\label{mass}
\end{equation}
In addition to these, there are an infinite sequence of poles
that reduces to a branch cut in the limit $H^2\to0$.
However, they have large imaginary parts and
do not give a dominant contribution to
the late-time behavior.

Now it is easy to see the late-time behavior of the Green function.
When $m_{\mathrm{eff}}^{2}/H^{2}<9/4$,
the contribution from the pole $p_+$ dominates.
The Green function after a sufficiently long lapse of time behaves as
\begin{equation}
 {\cal G}(x,x') \propto e^{(\sqrt{(9/4)H^2-m_{\mathrm{eff}}^2}-(3/2)H)t}.
\label{asym1}
\end{equation}
When $m_{\mathrm{eff}}^{2}/H^{2}>9/4$,
the contributions from both poles $p_\pm$ are equally important.
In this case, the asymptotic behavior of the Green function is given by
\begin{equation}
 {\cal G}(x,x') \propto e^{-(3/2)H t}
   \cos\left(
   \left[{m_{\mathrm{eff}}^2-{9\over 4}H^2}\right]^{1/2}t+\eta
\right),
\label{asym2}
\end{equation}
where $\eta$ is a real constant phase.
From the asymptotic behavior obtained in Eqs~(\ref{asym1}) and (\ref{asym2}),
we can conclude that the bulk
scalar field evaluated on the brane behaves as an effective
4-dimensional field with the mass squared $m_{\mathrm{eff}}^{2}$
given by Eq.~(\ref{mass})
after a sufficiently long period of de Sitter expansion.

\subsection{Effective 4-dimensional scalar field}
Let us now turn to the second order description of the system
where the backreaction to the geometry appears.
Since our solution guarantees the regularity of the geometry
in the bulk, we focus on the dynamics of the brane.
We consider the case when $|m^{2}|\ell^{2} \ll 1$
and $H^{2}\ell^{2} \ll 1$.

We note that from the late time behavior of the scalar field given by
Eqs.~(\ref{asym1}) and (\ref{asym2}),
the bulk scalar field satisfies the equation,
\begin{eqnarray}
\ddot\phi+3H\dot\phi+m^2_{\mathrm{eff}}\phi=0,
\end{eqnarray}
on the brane at late times.
On the other hand, the 5-dimensional field equation on the brane implies
\begin{eqnarray}
\ddot\phi+3H\dot\phi-\partial_r^2\phi+m^{2}\phi=0,
\end{eqnarray}
on the brane.
Using the effective mass given by Eq.(\ref{mass}), we obtain
\begin{eqnarray}
\partial_r^2\phi=(m^{2}-m^2_{\mathrm{eff}})\phi=m^2_{\mathrm{eff}}\phi
=-\ddot\phi-3H\dot\phi\,.
\label{dr2phi}
\end{eqnarray}
Inserting this to the integrand of Eq.~(\ref{weyl}),
we find
\begin{eqnarray}
E_{tt}&=&-{\kappa_5^{2}\over{2a^{4}}}\int^t a^{4}\dot{\phi}
(\ddot\phi + 2H\dot{\phi})\,dt
=-{\kappa_5^{2}\over{4a^{4}}}\int^t
{d\over dt}\left(a^{4}\dot{\phi}^2\right)\,dt
\nonumber\\
&=&-{\kappa_5^2\over4}\dot\phi^2,
\label{weyl2}
\end{eqnarray}
where we have neglected the integration constant term
($\propto a^{-4}$) that vanishes rapidly as time goes on.
Then we find the effective energy density on the brane,
given by Eq.~(\ref{rhoeff}), as
\begin{eqnarray}
\rho_{\mathrm{eff}}={\ell_0\over2}
\left({\dot{\phi}^{2} \over 2}+V(\phi)\right)-{E_{tt}\over\kappa_4^2}
={1\over 2}\dot\Phi^2+V_{\mathrm{eff}}(\Phi),
\end{eqnarray}
where
\begin{equation}
 \Phi=\sqrt{\ell_0}\,\phi\,\quad {\mathrm{and}}
\quad  V_{\mathrm{eff}}(\Phi)={\ell_0\over 2}V(\Phi/\sqrt{\ell_0})\,.
\label{cor2}
\end{equation}
Thus $\rho_{\mathrm{eff}}$ is given by an effective 4-dimensional
scalar field $\Phi$ with mass-squared $m^2_{\mathrm{eff}}=m^2/2$.
It is important to note that this is fully consistent with
the first order solution for the system where the bulk scalar field
is dominated by its zero mode.
To conclude, provided $H^2\ell^2\ll1$ and $|m^2|\ell^2\ll1$,
the effective dynamics of the Einstein-scalar system
on the brane is indistinguishable from
a 4-dimensional theory at the lowest order in
$H^2\ell^2$ and $|m^2|\ell^2$.

\section{Conclusion}

Based on the Randall-Sundrum type braneworld scenario, we
proposed a new model of the braneworld inflation in which the inflation
is caused solely by the dynamics of a 5-dimensional scalar field without
introducing an inflaton on the brane universe.
We noted that the scalar field we introduced in
this paper should presumably be a scalar field originating in gravity.
It is natural that the 5-dimensional action includes some scalar fields
of gravitational origin from the viewpoint of unified theories
in a yet higher dimensional spacetime.

Using the Green function given in Eq.~(\ref{calG2}), we derived the
late time behavior of the bulk scalar field under the assumptions
$|m^{2}|\ell^{2}\ll 1$ and $H^{2}\ell^{2} \ll 1$,
and showed that the bulk scalar field seen on the brane
behaves as a 4-dimensional effective scalar field with mass
$m_{\mathrm{eff}}=m/\sqrt{2}$, irrespective of initial
field configurations and of the value of $|m^{2}|/H^{2}$.
Moreover, we have examined the lowest order
backreaction to the geometry which starts at the
quadratic order in the amplitude of $\phi$.
We have found that the leading order backreaction to the geometry
is consistently represented by a 4-dimensional effective scalar field
$\Phi$ with the effective 4-dimensional mass $m_{\mathrm{eff}}$
mentioned above, where $\Phi$
is related to $\phi$ by a simple scaling (\ref{cor2}).
Thus this inflation model turns out to be a viable alternative
scenario of the early universe.

It is, however, the next order corrections in $H^2\ell^2$
and $m^2\ell^2$ that are of cosmological interest, since
these corrections are expected to give genuine braneworld
effects which can be used to test the scenario.
Furthermore, if we recall that the bulk scalar field is probably
of gravitational origin, it may be natural to expect $m^2$
to be of the curvature scale of the rest of compactified extra
dimensions which should be at least of the same order of
$\ell_0^{-2}$.  Since
$|m^2|\ell_0^2\lsim |m^2|\ell^2$, this implies $|m^2|\ell^2\gsim1$
for realistic models
(and which implies $H^2\ell^2\gsim1$ for inflation to occur).
It was shown in Ref.~\citen{Sago:2002gi} that the contribution
of the Kaluza-Klein modes are non-negligible when $|m^2|\ell^2\gg1$
(though small; of the order of 10\% at most)
and affect the brane dynamics even at sufficiently late times.
Thus it seems very important to study the case $|m^2|\ell^2\gg1$
in more details, or at least to clarify the effect of
the next order corrections in $|m^2|\ell^2$.
Investigations in this direction is in progress.

\section*{Acknowledgements}
We would like to thank T. Tanaka for fruitful collaborations
and discusssions on which a substantial part of this work is based.
We also would like to thank all the participants of
the YITP workshop on ``Braneworld --- Dynamics of spacetime with boundary"
for valuable discussions.
This work was supported in part by the Yamada Science Foundation
and by Monbukagaku-sho Grant-in-Aid for Scientific Research (S)
No.~14102004.


%

\end{document}